\documentclass{article}

\usepackage{arxiv}

\usepackage[utf8]{inputenc} 
\usepackage[T1]{fontenc}    
\usepackage{hyperref}       
\usepackage{url}            
\usepackage{booktabs}       
\usepackage{amsfonts}       
\usepackage{nicefrac}       
\usepackage{microtype}      
\usepackage{lipsum}		
\usepackage{graphicx}
\usepackage{natbib}
\usepackage{doi}

\title{Motivation, inclusivity, and realism should drive data science education}

\author{\textsuperscript{*} Savonen, Candace \\
	Fred Hutchinson Cancer Center\\ 
	Seattle, WA \\
	\And
    \hspace{1mm} Wright, Carrie \\
	Fred Hutchinson Cancer Center\\ 
	Seattle, WA \\
	\And
    \hspace{1mm} Hoffman, Ava M. \\
	Fred Hutchinson Cancer Center\\ 
	Seattle, WA \\
 	\And
    \hspace{1mm} Humphries, Elizabeth, M. \\
	Fred Hutchinson Cancer Center\\ 
	Seattle, WA \\
 	\And
    \hspace{1mm} Cox, Katherine E. L. \\
	Johns Hopkins Bloomberg School of Public Health \\ 
	Baltimore, MD \\
        \And
    \hspace{1mm}Tan, Frederick J. \\
	Carnegie Institution \\ 
	Baltimore, MD \\ 
         \And
     \hspace{1mm}Leek, Jeffrey T. \\
	Fred Hutchinson Cancer Center\\ 
	Seattle, WA \\
}

\renewcommand{\shorttitle}{Data Science Education}

\hypersetup{
pdftitle={Motivation, inclusivity, and realism should drive data science education},
pdfauthor={Candace Savonen, Carrie Wright, Ava M. Hoffman, Elizabeth M. Humphries, Katherine E. L. Cox, Frederick J. Tan, Jeffrey T. Leek},
pdfkeywords={education | data science | inclusivity | pedagogy },
}

\begin{document}
\maketitle

\title{Motivation, inclusivity, and realism should drive data science education}

\begin{abstract}

Data science education provides tremendous opportunities but remains inaccessible to many communities. Increasing the accessibility of data science to these communities not only benefits the individuals entering data science, but also increases the field's innovation and potential impact as a whole. Education is the most scalable solution to meet these needs, but many data science educators lack formal training in education. Our group has led education efforts for a variety of audiences: from professional scientists to high school students to lay audiences. These experiences have helped form our teaching philosophy which we have summarized into three main ideals: 1) motivation, 2) inclusivity, and 3) realism. To put these ideals better into practice, we also aim to iteratively update our teaching approaches and curriculum as we find ways to better reach these ideals. In this manuscript we discuss these ideals as well practical ideas for how to implement these philosophies in the classroom. 

\end{abstract}

\section*{Keywords } education | data science | inclusivity | pedagogy 

\let\thefootnote\relax\footnotetext{ * Corresponding author: \href{mailto:csavonen@fredhutch.org}{csavonen@fredhutch.org}}

\clearpage
\pagestyle{fancy}

\section{Introduction}

Data science is booming and many fields, including computational biology, have rapidly evolving data science needs \citep{labor_stats, przybyla_should_2020}. For these needs to be met, scalable education efforts need to be supported \citep{demasi_ad_2020}. Data science classes are often taught by practicing data scientists who have taken on a teaching role. This means they may have an idea of the changing landscape of data science but may lack experience in education. This phenomenon has been discussed in academia more generally, and it may be especially relevant to data science teaching. \citep{flaherty_required_nodate, robinson_teaching_2013, stenhaug_teaching_2019}. Data science experts can be very passionate about teaching the next generation but educators also need training in education methods for writing curriculum, lecturing, creating assessments, constructing learning objectives, or the many other duties that accompany teaching roles \citep{flaherty_required_nodate, robinson_teaching_2013}.  

Talent for data science careers is equally distributed, but opportunity for such careers is not \citep{Janah2014, datatrail, genomic_data_science_community_network_diversifying_2022, hall_mark_academic_2013}. Educational barriers in STEM often begin as early as elementary school \citep{fuller_2021}. However, access to education and resources as well as efforts to raise awareness have the potential to reverse this pattern \citep{canner_2017}. As a next step, helping empower educators in best practices can help make opportunities for data science careers more equitable. Data science educators must tailor their teaching methods to equip students with relevant skills, while also contextualizing the material based on the students' interests and backgrounds. \citep{hazzan_2023_what_is_data_science}. Furthermore, democratizing this knowledge to enhance more diverse representation in data-driven fields holds the potential to increase innovation \citep{hofstra_diversityinnovation_2020} while avoiding harms \citep{de_hond_2022}.

Data science education programs have become increasingly popular. Organizations like The Carpentries and Dataquest, fast.ai, as well as massive open online courses (MOOCs) and formal Master's programs and certificates in data science indicate the great demand for these materials \cite{software_carpentries, noauthor_dataquest_nodate, noauthor_fastai_nodate}. MOOCs, while helpful, tend to primarily benefit highly educated individuals \citep{emanuel_moocs_2013}. Students often need instructors who understand their needs and are willing to work with and build relationships with them. While data science instructors can access these excellent and inspiring training materials directly, educators are often in need of more guidance. For example, The Carpentries has an \href{https://preview.carpentries.org/instructor-training/}{instructor training course} that covers teaching approaches and skills that emphasize motivating, inclusive, and accessible practices \citep{software_carpentries}. 
For implementation at the high school age level, the Introduction to Data Science curriculum from UCLA \href{https://curriculum.idsucla.org/}{has materials for training teachers} \citep{Gould2014}. 

In this opinion article, we combine advice from existing resources with our own education experiences for instructors who plan to take the next step in designing and implementing data science educational content.  Our lab is involved in a number of data science education efforts, including training high school students, graduate students, postdocs, researchers, and university faculty, and these experiences have taught us a great deal about data science teaching that may be useful for others’ education efforts \citep{genomic_data_science_community_network_diversifying_2022, wright_open_2023, kross_end-user_2019}. We will discuss the overarching lessons we have learned as well as practical tips for how these lessons can be applied in the classroom. The deeper learning occurs when we apply or create using what we've learned \citep{bloom_taxonomy_1956}. We encourage data science educators to use any advice from this discussion that best fits their classroom and audiences. We will also strive to better apply these ideals in our own teaching as we continue to learn. For summaries and lists of resources from this manuscript, see the Supplementary Materials section. 
  
\subsection{Teaching Ideals}

\begin{figure}
\centering
\includegraphics[width=.5\linewidth]{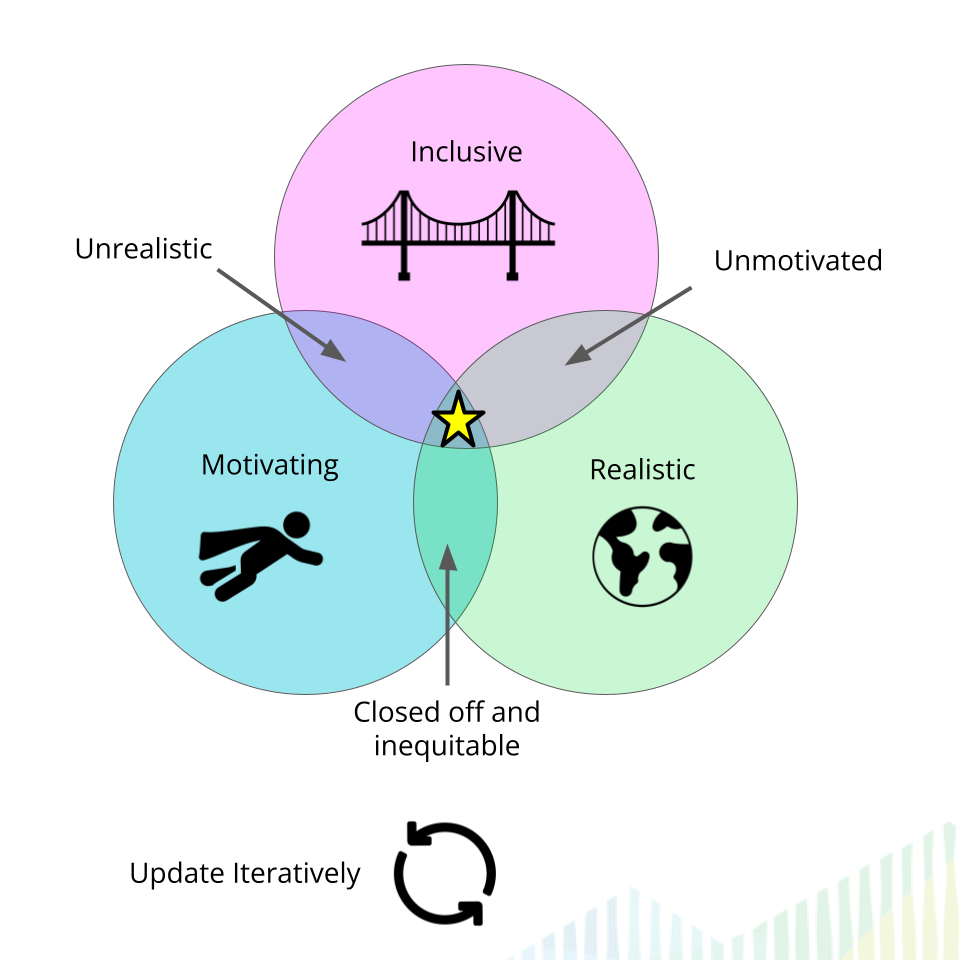}
\caption{A Venn diagram illustrating our teaching philosophy. Ideally, data science teaching should be motivating, inclusive and realistic. If any of these ideals are lacking, our teaching effectiveness suffers, which is why the fourth ideal includes iteratively updating our curriculum and approaches as we find better ways to meet these ideals.}
\label{fig:philosophy}
\end{figure}

The lessons we have learned inform our teaching philosophy, summarized into these main ideals (See Figure~\ref{fig:philosophy}). 

\begin{itemize}
\item \textbf{Motivation}: Aspiring data scientists face many demoralizing steep learning curves. As others have noted \citep{software_carpentries}, motivation is key for learners to persist and succeed despite these challenging learning curves of data science.
\item \textbf{Inclusivity}: Diversity is lacking in data science education \citep{general_assembly}. Making data science more inclusive is not only the right thing to do, but improves innovation and understanding \citep{hofstra_diversityinnovation_2020, tomasev_2020, gaynor_2022}. Data science suffers when there is inequitable entry into the field. 
\item \textbf{Realism}: The best learning approaches are those that attempt to prepare students for "real life" as much as possible \citep{meyers_how_2009}. We strive to have our curriculum be hands-on and interactive in ways that reflect what our students will be doing as data scientists in their communities and outside the classroom. 
\item \textbf{Iteratively update}: Students are not be the only ones learning. Data science is a fast changing field. Not only do we need to keep up with data science as a field to best prepare our students, but we also need to learn best teaching practices from education research \citep{schwab_mccoy_2021}.
\end{itemize}

Our teaching philosophy depends on all of these ideals together. As the diagram illustrates, when one ideal is lacking, our teaching is less effective. Without motivation, students will struggle to persevere through challenging material. Without realism, our students will not be adequately prepared for careers as data scientists. Inclusivity makes these ideals more democratic. Here, we will discuss practical implementation of motivating, realistic, and inclusive data science teaching.

\subsection{Education Projects}

The ideals and practical advice in this paper have come from our collected teaching experience in these various projects we will highlight. 

\subsubsection{DataTrail}
\href{https://www.datatrail.org/}{DataTrail} is a 14-week paid educational initiative that promotes inclusivity in data science education for young adults, high school graduates, and GED recipients. We provide financial, social, and academic support, including weekly check-ins, tutoring, and internships for graduates. Our program goes beyond programming to prepare students for real-world situations. We continue to learn lessons about how our teaching has helped our students in their internships following graduation \citep{kross_end-user_2019}. 

\subsubsection{GDSCN}

The \href{https://www.gdscn.org}{Genomic Data Science Community Network} works to broaden participation in genomics research by supporting researchers, educators, and students from diverse institutions. These institutions include community colleges, Historically Black Colleges and Universities, Hispanic-Serving Institutions, and Tribal Colleges and Universities. Our vision is one where participation is not limited by an institution's scientific clout, resources, geographic location, or infrastructure. These institutions play a critical role in educating underrepresented students \citep{li_2007}. We work with and support faculty from these institutions as they address systemic bottlenecks.

\subsubsection{ITCR Training Network}

The \href{https://www.itcrtraining.org/}{ITCR Training Network} supports cancer research by equipping users and developers of cancer informatics tools with training opportunities. These audiences are professional learners with advanced degrees. They often have specific project goals that they are looking to apply data science skills to. We attempt to equip the users of cancer informatics tools with data science foundations they need. We also have other training opportunities for the developers of cancer informatics with skills and principles they can use to increase the usability of their data science tools.  

\subsubsection{Open Case Studies}

The \href{https://www.opencasestudies.org/}{Open Case Studies project} provides an archive of experiential data analysis guides that covers analyses using real-world data from start to finish \citep{wright_open_2023}. There are over 10 case studies that are currently focused on utilizing methods in R, statistics, data science, and public health to evaluate timely and relevant public health questions. These case studies are intended to be used by instructors to assist with teaching courses, as supplemental resources for courses, or as standalone resources for learners. They are aimed for undergraduate and graduate learners, but also have material appropriate for teaching high school students. The case studies help showcase the data science process and demonstrate the decisions involved. 

\section{Motivation} 

A common barrier for individuals entering data science is motivation \citep{woolley_motivating_2022} to overcome the steep learning curves involved in becoming a data scientist -- namely learning programming and statistics. For many students, having a computer print "Hello world!" isn't enough to sustain their interest through the trial and error of background programming and statistics necessary for more inspiring data visualizations or interactive apps. Motivation is the fuel that is needed for successfully overcoming these learning curves \citep{software_carpentries}. Budding potential data scientists need to be made aware that frustration, failure, and mistakes are normal and do not indicate that they are unsuited for the field! 

The best motivation for data science is usually not an deep interest in programming or statistics, rather it is a data related question or problem that a budding data scientist cares about. This is why data science is such a broad and pervasive field. Data scientists may arise from a variety of different questions, problems and contexts. The best motivation for becoming a data scientist is to have a problem that has a quantitative question behind it.

Its also worth noting that some students may never truly be interested in the programming side of data science as a career and that is also okay. It is also our goal to increase data literacy so that students who pursue other interests walk away with useful interdisciplinary skills.

\subsection{Confidence and repetition are more important than talent}

Learning data science is hard. This means learners tend to believe that they are “just not good at it" and that this is a fixed intrinsic quality they possess. These beliefs can affect not only a student’s confidence but perhaps worse, a teacher’s behavior \citep{makarova_gender_2019}. We have a tendency to attribute innate talent to one's success instead of realizing that it is often due to dedication and practice \citep{chambliss_mundanity_1989}. In some ways, learning to program is no different from learning a spoken or written language. Practice and mistakes build fluency. But many students become disheartened if they do not learn programming right away.  Instead, we want to encourage a "growth" mindset \citep{software_carpentries, growth_mindset}. A growth mindset emphasizes that success is not about where you started, but the idea that you can continue to improve your skills if you continue practicing. 

Through the DataTrail program, we have witnessed students become disheartened as data science skills become difficult to learn. To better encourage our students, we reorganized our curriculum to get to the "fun" things sooner. Previously, we didn't teach data visualization until the last quarter of the course. We realized that making visuals is something students generally enjoy and can boost confidence. Students often learn best with utilizing different modalities (images and non-images for example) \citep{clark_dual_1991}. So now, we have students create visuals in the very first project (with some help of code we have pre-written for them). Not only has this helped motivate our students, but it has the added benefit of giving us more insight into what students are writing and trying with their code. 

\subsubsection{Practical ideas for boosting confidence}
\begin{itemize}
    \item \textbf{Get to the magic:} Have your students see a glimpse of the "magic" early on. Images, plots and interactive pieces of code and apps are the most fun. You can use "cooking show" magic and provide your students with something 99\% processed but then have them be able to customize or fill in the last bit of code.
    \item \textbf{Emphasize the "yet":} Encourage a growth mindset by assuring the student that with practice and perseverance they will learn this even if they haven't learned it "yet". \citep{software_carpentries}.
    \item \textbf{Validate trickiness:} Try to stay away from using phrases like “it’s easy”. Although you might be attempting to impart confidence, it may make students feel inadequate if they don’t also find it easy. Validating the challenges in data science can be reassuring and help renew a student’s confidence. 
    \item \textbf{Assure togetherness:} Sometimes frustration regarding a problem can feel lonely and impossible. Assuring the student that you are there to help can be a powerful learning tool. 
    \item \textbf{Celebrate the wins!:} No matter how small the win, celebrate what your student has accomplished. This can mean congratulating them, but also may mean encouraging them to share their success with their peers. 
    \item \textbf{Do not compare:} Be careful to not compare your students' skills with your own or with other students. Emphasize the idea that everyone has a different starting points and aptitudes. What matters is not where you are today, but that everyone continues to learn.  
    \item \textbf{Have madlibs code:} By madlibs code, we mean code that is mostly written but has the student fill in the blanks. This allows the student to see the control they have over code without requiring them to write a whole project from scratch before they are ready to do so. 
\end{itemize}

\subsection{Mistakes are good and should be expected}

Feeling a bit uncomfortable and making mistakes is an important part of learning. In fact, we remember better when we make mistakes \citep{cyr_mistakes_2015}. We also want everyone to feel comfortable with what they do not know. This starts with the instructor! It is incredibly valuable for a student to see how their instructor works through a problem or goes about finding a solution. It can be tempting to feel like you need to be a font of wisdom and it can feel like it undermines your authority if you let students see you struggle through a solution. However, it is just as important to model how to tackle the unknown as it is to demonstrate your knowledge and mastery. It can be very helpful to model the process it takes from not knowing something and making mistakes and iteratively figuring it out. Evidence shows that people who expect discomfort and challenge are more likely to overcome those challenges and perceive their accomplishment more positively \citep{woolley_motivating_2022}. 

In the DataTrail program, before we begin to code, we attempt to help manage our students' expectations about programming by discussing the role mistakes have in data science. We have \href{https://datatrail-jhu.github.io/DataTrail/how-to-learn.html}{a chapter about how to learn in data science} with a strong emphasis that data science involves a lot of questions and failure and that is good!

\subsubsection{Practical ideas for encouraging mistakes}
\begin{itemize}
    \item \textbf{Talk about mistakes you've made:} It can be helpful if as the educator you also tell of times you've made mistakes in your work. It can help students start to reverse the idea that mistakes should be hidden but instead that mistakes are normal! This is something that we've embraced from the \href{https://builtin.com/data-science/data-mistakes}{Data Mishaps Night} \citep{datamishaps}.
    \item \textbf{Model making mistakes:} Although you don't need to purposely make mistakes in live coding, if you do see yourself starting to make a mistake, maybe don't stop it right away. Try to let students catch your mistake instead of pointing it out to them. When they do point it out to you, be sure to be happy about the fact that as a team, you have caught the mistake. 
    \item \textbf{Say "I don't know":} If a student asks a question that you are unsure of, say "I don't know" proudly. Use this as an opportunity to demonstrate that not knowing something is expected. The class can also take the opportunity to look something up together. If the question is less relevant to the whole class, let the student know you will look into it and get back to them or you and the student can look into it together later.  
    \item \textbf{Encourage iteration:} Reaffirm the idea that drafts are okay. Blank pages are harder to work from than pages full of mistakes. Rather than striving for perfection on the first try, emphasize the idea that we can use version control and return to this code later to better polish it. 
    \item \textbf{Normalize questions:} Instead of “are there any questions?”, ask “what questions do people have?”. This small wording change can help lower the intimidation factor by implying questions are to be expected. 
\end{itemize}

\subsection{Atmosphere boosting to encourage learning}

For people to learn, they need to feel comfortable. A great way to make people feel comfortable is by being more informal and showing it is okay to be silly. Our education group encourages using silliness as a teaching tool. Research shows that humor can increase motivation, learning, and perception of authentic teaching \citep{johnson_examination_2017, banas_review_2011, wanzer_explanation_2010}. Silliness defuses frustration and lowers our anxieties about a topic \citep{woolley_motivating_2022}. We prioritize being silly over seeming important and solemn. Silliness is an atmosphere booster and can help educators seem more approachable. Being silly means: minimizing use of jargon, making it okay to laugh at ourselves, and trying to connect with people. Being comfortable with yourself also makes you a good role model. Learners, particularly those who are underrepresented, benefit when they see their instructors are real people \citep{reupert_2009, Pacansky-Brock_2020}. Your own brand of in-class examples or metaphors are often the ones that stick with students the longest. 

\subsubsection{Practical ideas for being silly}
\begin{itemize}
    \item \textbf{Study silly data:} Data science doesn't always have to be about life changing questions. Sometimes it can be a lot of fun to analyze datasets about \href{https://www.kaggle.com/datasets/rounakbanik/the-movies-dataset}{movies}, \href{https://datahub.io/five-thirty-eight/candy-power-ranking}{the best halloween candies}, and \href{https://www.kaggle.com/datasets/josephvm/bigfoot-sightings-data}{bigfoot sightings}.
    \item \textbf{Use GIFs and cartoons:} Cartoons and GIFs can be mood boosters and can even make salient points that will stick with your students after class is over. 
    \item \textbf{Use fun data examples:} People like movies and pop culture. So long as the data is appropriate, it can be fun to use data examples that has material that people enjoy.
    \item \textbf{Use silly icebreakers:} The more the classroom feels comfortable with each other, the more they will be ready to learn and participate. Let the class know its okay to be silly by asking a silly question.
    \item \textbf{Take snack breaks:} Snack breaks or other kinds of breaks don't need to be silly per se, but being silly during breaks are important to help people stay refreshed and ready to learn.
\end{itemize}

\section{Inclusivity} 

Data science suffers from a lack of diverse perspectives. It is disproportionately white, upper class, and male; concentrated in select geographical locations; harder to access by persons with disabilities; and challenging for first-generation students \citep{datasciencedemo}. Some of the largest biomedical data science institutions are located in areas with low income mobility \citep{datatrail} and likely have contributed to lack of opportunities \citep{dawkins_2023}. However, income mobility can be mitigated by education \citep{chetty_association_2016}. Inclusive and diverse research and science benefits individuals and is also more innovative \citep{hofstra_diversityinnovation_2020}. We therefore prioritize the promotion of Inclusion, Diversity, Anti-Racism, and Equity (IDARE) through our curriculum and in our classrooms. Everyone learns more and is more productive when we emphasize inclusive practices and continue to look for opportunities for learning and growth \cite{NIH2023, puritty_without_2017}. 

\subsection{Underrepresented communities face barriers}
\begin{figure}
\centering
\includegraphics[width=.9\linewidth]{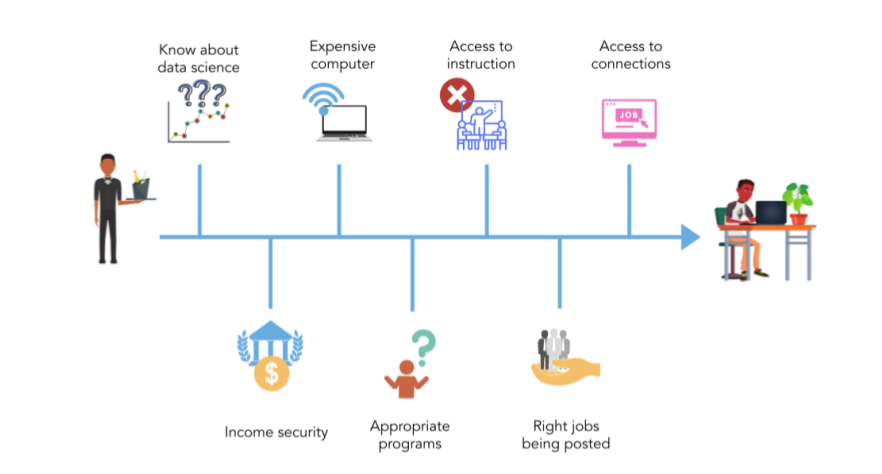}
\caption{This figure represents the barriers that many members of underrepresented and underserved communities face when entering the field of data science. Some of the barriers to entering data science include: Knowing about data science as a career option, income security, access to an expensive computer, access to instruction, knowledge of the appropriate education programs, having connections to the industry and having the right jobs being posted, and knowing how to look for the appropriate job openings.}
\label{fig:barriers}
\end{figure}

Despite increased awareness, large disparities in funding and support persist across primary and secondary education. These disparities have generally been at the expense of underrepresented groups such as students from Black and Indigenous communities. Primary schools with greater resources might be able to introduce data science at much earlier ages, enhancing students’ ability to pick it up and understand it \citep{morgan_science_2016, lee_call_2021}. By contrast, students from under-resourced schools might have few opportunities to learn about data science. Students might also encounter barriers when gaining access to a computer and/or internet, securing childcare or time off work, and breaking into job networks (See Figure~\ref{fig:barriers}). These systemic disparities in education mean that learners from underrepresented groups may need more support\citep{genomic_data_science_community_network_diversifying_2022}. 

\subsection{You are not your student}

In user experience design, there is a saying that “you are not your user" \cite{Savonen2021}. In data science education, we could just as easily say “you are not your student”. Help your students find their passion in data science realizing it might differ from yours. Students will be more motivated to persevere through the challenges of data science (such as learning programming syntax and troubleshooting confusing errors) if they are fueled by a curiosity or passion behind their data science project. Instructors should ask students about their interests and what they want to do with their career to encourage exploration that aligns with their interests \citep{genomic_data_science_community_network_diversifying_2022}

Also realize that students from underrepresented backgrounds likely have talents that are unrecognized or unrewarded by our traditional educational systems. Students have diverse talents and likely have awareness of important problems that might be missed by instructors. Try to emphasize in the early stages of a data education about how to bring these skills to the surface. Data science education needs to focus on more than just the technical skillsets. 

If you have been coding for some time, you have likely forgotten how frustrating it was for you when you first started. Not only should you have empathy for your students as they approach this steep learning curve, but you also should remember that your students may have very different experiences and backgrounds than yours. Be aware not everyone "gets it" in the same way you do. Going back through the material as if you are witnessing it for the first time with empathy for students who are completely unfamiliar and noticing what information you might take for granted as common knowledge can help. In addition, teaching a topic that you have just learned, can also help. 

\subsubsection{Practical ideas decreasing barriers}
\begin{itemize}
    \item \textbf{Encourage or require office hours:} Many students believe office hours are not relevant to them \citep{abdul_2019}. Encourage your students to drop by and introduce themselves any time, not only when they are stuck.
    \item \textbf{Survey your students:} In-class anonymous surveys can answer questions like "why are you taking this class?" and "what career fields interest you?", which can serve as a starting point for conversation.
    \item \textbf{Provide Mentorship:} Young data scientists need support, particularly those from disadvantaged backgrounds. When possible, try to connect learners to supportive mentors who have time and understanding to devote to the learner. Ideally a mentor can be someone of a similar background to help encourage the learner through shared experience and understanding, but any form of mentorship is still beneficial.  
    \item \textbf{Don't require people to buy expensive things:} Income insecurity can be a massive barrier to entry into the field of data science but it doesn't have to be. When possible, pursue cloud-based computing resources for your students to use \citep{kumar_2017}. This will allow them to run more computationally costly analyses on nearly any machine, including relatively inexpensive computers like Chromebooks \citep{datatrail}.
\end{itemize}

In the DataTrail program, we attempt to mitigate many of these barriers. We provide individuals in the program with inexpensive Chromebooks that allow students to use cloud programming platforms to conduct their analyses. We pay individuals to participate in our program to help mitigate the issue of income insecurity. We also have partnered with non-profit hiring partners to place graduates of the program in internships. Even with our program's social and financial supports, it is difficult for individuals to overcome the institutional and systemic biases that have been rooted in our society, but we hope that these supports give improved access to the individuals in our program. We encourage other data science program administrators and instructors to attempt to add financial and social support for underrepresented individuals whenever possible.

\subsection{Practical ideas for inclusive classrooms}

\begin{itemize}
    \item \textbf{Frequently take the temperature of the room:} Silence and pauses may feel awkward but they are critical to good teaching. Allow students to have time to think. Another useful tool is using sticky notes to keep track of whether students who are actively coding need help or are doing okay. People who are doing well can put up a green post it, while people who need help with something can put up a different color post-it \citep{software_carpentries}. Try to create an atmosphere that helps to decrease the intimidation factor of asking questions. Additionally, tools like \href{https://www.slido.com/}{Slido} can help you collect interactive responses from your students from their smartphones or computers \citep{slido}.
    \item \textbf{Explain things in multiple different ways:} Perhaps you understand things well using a particular analogy. But that analogy may not resonate with all your students. Try to think outside of the box and explain things in multiple different ways.
    \item \textbf{Do not assume everyone knows the basics:} Err on the side of explaining the most fundamental piece of knowledge. Less experienced students will be less likely to get lost when the curriculum advances, but more experienced students may overestimate how well they know something. Everyone benefits from starting off on the same page with the basics. 
    \item \textbf{Use inclusive language:} Refer to guidelines for creating inclusive communities \citep{lee_2016_inclusive_community}. Educators can unknowingly use phrases that reinforce stereotypes or perpetuate gaps in the STEM fields \citep{software_carpentries}. This also means that as an educator you should always be ready to be corrected and change course should a student share with you how they could be better accommodated. 
    \item \textbf{Look for ways to improve the accessibility of your classroom:} This includes simple things like making sure your curriculum can be read by a screen reader and testing your curriculum for color vision compatibility with tools like ColorOracle \citep{color_oracle}. See our Supplemental information for a longer list of accessibility items to consider.
    \item \textbf{Be aware of implicit biases and stereotype threat:} Though behaviors with implicit bias are by definition unintentional, they can be very harmful all the same \citep{APA_implicit_2023}. See our Supplemental information for resources and classes to take to combat implicit bias, stereotype threat and related issues. 
\end{itemize}

\section{Realism}

As educators, a key goal is to prepare students to pursue their interests in their chosen careers. We do learners a disservice if we do not adequately teach them the skills they will need. Instructors must take note of the data science needs, whether within large companies or smaller community non-profits, and teach those skills. This includes collaborative "soft" skills, like giving/receiving feedback and code review. This also means using real data and real workflows \citep{wright_open_2023} as well as incorporating data ethics and domain specific contexts \citep{oliver_2021}. 

Being realistic also means utilizing the ideas of "just in time teaching." \citep{Novak2023}. We want to bring the concepts we discuss into real life scenarios as quickly as possible. Anyone who has been taught a complicated board game has felt the importance of just in time teaching. It is often overwhelming to be told a long list of rules and concepts that mean nothing to you before you have even begun to apply anything. Similarly, it is highly stressful to be told to remember things and “it will make sense later.” A more effective teaching tool to tell your students something and then directly apply it in an activity. Not only does this better align with a deeper level of application on the Bloom's taxonomy educational objectives, it also avoids oversaturation \citep{bloom_taxonomy_1956}. Application and practice are key. 

Practically, this means introducing a concept and following up quickly with live coding,  concluding with hands-on practice by the learners themselves. Live coding is more useful than long lectures about concepts and helps ease learners into trying the code themselves \citep{nederbragt_ten_2020}. Perhaps the most useful bit of live coding is when you employ strategies discussed or accidentally write errors yourself \citep{shapiro_teaching}! If you make an error while live coding, it allows students to see that everyone mistakes and provides an opportunity to demonstrate how you might go about investigating an error.

Providing concepts using a variety of real datasets and asking real data science questions relating to a diversity of learners can also be extremely motivating and helpful for students, especially when they try applying what they learned to other contexts \citep{podschuweit_composition-effects_2018}. Not all applications of teaching in “context” are useful. We have found that this needs to be done in a careful manner that does not overwhelm the learners as well as with intention to describe the data analysis process. Focusing on simple examples with fewer datasets helps to first introduce topics, particularly for beginners. This can be followed up with more examples and discussion of when certain methods need to be used for different types of data and analyses. 

Soft skills like communication are critical but often overlooked in curricula. Students may need explicit training in areas such as asking questions about a project, giving effective presentations, writing professional emails, providing feedback to colleagues, or participating in meetings.  In earlier iterations of DataTrail we did not emphasize these skills as we did not realize how much training  our  decades of professional experience  had given us. We now include a unit on Communication (both formal and informal) and a unit on Career Development as part of the DataTrail curriculum, and we continue to refining our approach. 

\begin{figure}
\centering
\includegraphics[width=.8\linewidth]{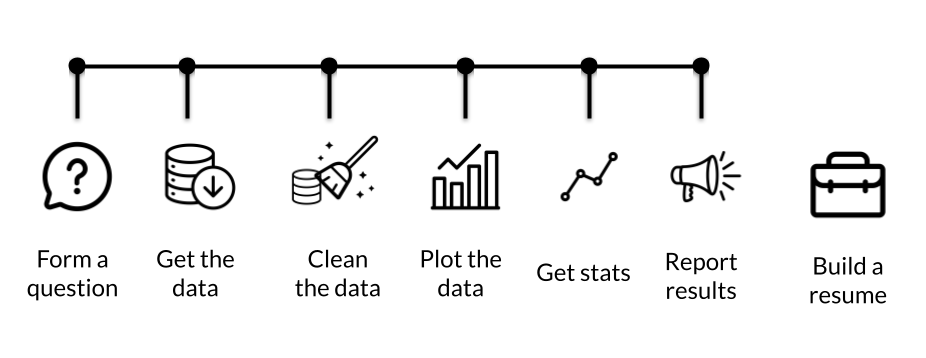}
\caption{Our DataTrail curriculum is now structured to reflect the typical steps performed in a data analysis. Our earliest chapters cover how to form a data science question, and then the following chapters take the students through how to get data, clean data, create visualizations, collect statistics, and share those results.}
\label{fig:curriculum}
\end{figure}

Additionally, we realized that we need to have our students learn in a project-based manner that better reflects "real life" data. We restructured our curriculum to have more projects earlier so that the whole curriculum parallels the steps that are taken in a "typical" data science workflow. Our projects start out heavily scaffolded, with a lot of the necessary code for a project written in so students only have to fill in minor steps, but with each chapter we leave more and more of the data science process to the students to determine for themselves (See Figure~\ref{fig:curriculum}).  

\subsection{Data science involves teamwork}

Data science is best done as a part of a team. It involves being relatively skilled at many different fields: computer science, statistics, writing, web design, programming, etc. Likely one person will not be an expert at all of these things, which is why teamwork is vital to good data science. Help your students realize that there's no such thing as a "lone genius". In real life teamwork not only helps us all learn, it also creates better end products \citep{parker_opinionated_2017}. Increasing the diversity of data science teams can increase the diversity of perspectives and potential solutions! 

\subsubsection{Practical ideas for promoting teamwork}
\begin{itemize}
    \item \textbf{Use pair programming:} Have designated time to have students practice paired programming. Or have optional or required time that students can pair program with you or other tutors.
    \item \textbf{Cover code review:} Explicitly cover techniques for how to conduct formal code review and why it is important \citep{Savonen2022} \citep{bacchelli2013}.
    \item \textbf{Highlight coding communities:} Introduce your students to online or in-person coding communities such as R-Ladies, StackOverflow, etc. \citep{stackoverflow}.
\end{itemize}

\subsubsection{Data science requires being flexibly prescriptive}

Data science projects can be quite varied and decisions can be stressful. Being flexibly prescriptive means giving specific instructions and making choices for them now, while preparing students for deviating from these choices later depending on what their project calls for. In practice, this means showing students one way to do something while letting them know about the existence of other relevant ways to do it that they may encounter. Data science, and code in particular, can have unlimited numbers of solutions to reach the same endpoint. Ultimately, some decisions are made based on what is comfortable to the data scientists working on the project, while other decisions may be based on what the project calls for. Instructors should give students something to start with but also acknowledge that they might use different methods in the future and that is okay. 

\subsubsection{Practical ideas for being flexibly prescriptive}
\begin{itemize}
    \item \textbf{Be aware of the stage of your audience:} What concepts do your students understand well? What concepts overwhelm them? If you are at an early stage of the process where students are attempting to grapple with a lot of information at once, do not bring up alternatives.
    \item \textbf{Acknowledge the existence of alternatives} If learners are likely to encounter common alternatives for particular methods in the real world, be upfront about this. This does not mean that learners necessarily need to dive into these alternatives, but simply noting the names of such methods can enable learners to recognize them in the future. 
    \item \textbf{Many solutions to the same endpoint:} Reinforce the idea that there may be a multitude of ways in code to reach the same endpoint. The priorities should be that the code works and is relatively readable.  This can tie in well with practicing code review, which lets them see and evaluate approaches taken by others.
\end{itemize}

\subsubsection{Keep it simple}

It is often more difficult to figure out what to skip in the curriculum as opposed to what to discuss. The best lessons are based on well-structured learning objectives with relatively narrow scope. It is often better for students to walk away understanding a few things well, rather than many things shallowly or not at all.  As an enthusiastic teacher who is very knowledgeable, you may have an impulse to teach everything to your students all at once. As scientists, many of us feel an impulse to say “well actually” or “technically” and give more nuance than is needed at a particular stage. It is the educator's job to curb these impulses and try to remain as simple and focused as possible. For example, it is not exactly correct to say that Docker is a virtual machine, but when explaining what Docker is, it can be helpful to explain it as "a computer that you run on your computer to ensure the same specs as another person". At a later point in time, when the student is ready, they might be able to replace this partially incorrect concept with a more nuanced one. Understanding nuance also means understanding what learners need or want to know and respecting that they are also busy with other things in their life.

In the DataTrail program, our original curriculum had a disproportionately deep level of coverage of statistical theory. We found that this did not match our students' career goals or background knowledge and was overwhelming. We restructured our lessons about statistics away from high level theoretical concepts to instead be based in practical examples, focusing on how to make decisions about which tests to use and how to interpret results in plain language. 

\begin{figure}
\centering
\includegraphics[width=.7\linewidth]{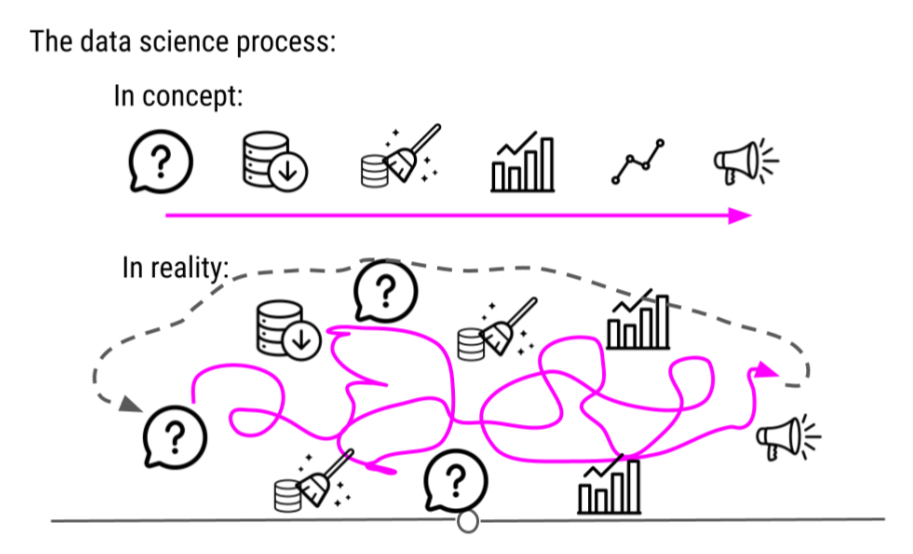}
\caption{This is an example image from our DataTrail course demonstrating how we sometimes use oversimplification as a tool in our curriculum. In the earlier chapters of the course, we tell students that the data science process is linear. In the later half of the course, we add on to this concept by letting them know that actually, data science is rarely a stepwise, linear process, but instead a process that involves a number of side investigations that may or may not lead to dead ends.}
\label{fig:simplify}
\end{figure}

Keeping it simple is also advice relative to the stage of your students. For example, in the DataTrail program, we initially keep it simple by introducing data science as a linear and step wise process. Later in the course, we expand on this to explain to our students that actually, the data science process is rarely linear and usually involves a lot of side investigations, dead ends, and sometimes starting from scratch entirely (See Figure~\ref{fig:simplify}).

\subsubsection{Practical ideas for keeping it simple}
\begin{itemize}
    \item \textbf{Stop yourself:} Interrogate why you might give a complex answer. Is it because you want people to know you are knowledgeable? Are you very enthusiastic about the material? Remember to focus on the learner. Too much nuance or focusing on exceptions to rules will likely be a disservice to their learning experience. 
    \item \textbf{Chunk it out:} In code outside of the classroom, you may try to reduce the number of lines and put similar steps together. However, in the classroom, it can be beneficial for you to break down each step separately. This may look like making one chunk of code into multiple separate steps that you walk through with the students. You should also explain and encourage students how to chunk out code for their own troubleshooting technique.  
    \item \textbf{Keep it practical:} You likely have a lot more information about a topic than you need to share. Ask yourself what practical information would your students need to know in a "real world" data science project? For many topics, they won't need to know deep history or the ins and outs of each parameter of a function. We need to be selective about when history of something aids to understanding and when it does not. 
    \item \textbf{Make it skimmable:} You may notice we use lists and bold type in this paper to highlight main points. Respect that your students are busy and your class is not the only thing they have going on. What's the most efficient way for you to communicate this (either in print or verbally)? 
    \item \textbf{Link it out:} If you have additional information for the particularly curious student, feel free to share it, but don't make it central. Add links or a collapsible menu where students can find more information, but don't use time in a lecture to cover it. 
\end{itemize}

\subsubsection{Encourage curiosity}

Some of the best investigative data science starts off with a “that's weird". This concept applies to multiple scenarios. Science means chasing the "why" of an unexpected result. It may be the truth is weird, or there could be a mistake in the data handling. The only way to find out is to poke around at each step. Similarly, the best way to understand how code works is to take someone else’s code and change it. You will inevitably break the code, but as you break it, you will learn what each part of the code does. To find out how code works, investigate it piece by piece while examining the output of each part. Trying each piece of code interactively can help you build together what a longer line of code is actually doing. 

\subsubsection{Practical ideas for teaching students to encourage curiosity}
\begin{itemize}
    \item \textbf{Pause and think:} Sometimes in an effort to complete a project quickly we can move too quickly and miss a critical clue in the data. Pauses are effective tools for thinking effectively about a project and what you are seeing. Encourage students that they do not have to answer questions right away. They can walk away, think about it for minutes or days and come back to it. 
    \item \textbf{Show real examples:} Real data have weirdness. Your curriculum should include an example of real data weirdness and how someone found that weirdness. What functions were used? What aspects of the data were the first red flags that the person who did the data analysis followed? Tell the story about how we found out this weird thing about this real data. 
    \item \textbf{Give them an investigative tool belt:} Give your students a set of strategies they can use to investigate weirdness. What functions or tests can they use to interrogate a piece of data or weirdness in a package? Where can they go to find more information? Demonstrate Googling, StackOverflow, and package documentation as investigative tools. 
    \item \textbf{Model investigative data science:} In a live coding or pair programming session, encourage your students to look for abnormalities. Ask them questions about what they think about the results or what we might want to look out for. Model checking your data after each step.
    \item \textbf{Leave in the side journeys:} Often, complete analyses involve several side explorations and dead ends. Although we often want to show a polished data science story, sometimes it can be beneficial to briefly demonstrate your development process and the side journeys it took to get there.
\end{itemize}

\subsubsection{Communication is critical}

The most ground-breaking data science project is not worth anything if its results and importance cannot be communicated to others. While programming abilities are important, communication, documentation,  and other professional skills are perhaps even more important -- and unfortunately often harder to teach than programming. These skills have generally not been taught directly at educational institutions but historically have been passively learned through being in the workplace. But given that data science today is commonly remote work, it can be challenging for learners to build their communication and task management skills. These communication skills will be critical for learners success in the data science environment. Data science communication can be summarized by a few different aspects: 

Simple analyses with well communicated results are always better than overly complicated analyses that are poorly communicated. Data scientists translate numbers into stories that we can act on. Presenting results is just that -- telling stories about the data science project journey. This also includes recognizing misleading visualizations and avoiding using them for communication. In code and results, we should be writing down our thoughts as we are developing analyses. For more about good documentation you can see our course \citep{Savonen2021}. The example code that you use to teach should self-explanatory. Documentation must make sense to whoever will be using the code, whether this is the instructors, team members, or students themselves.

Communication in data science is also critical as a part of a team. Every data scientist gets stuck at some point and needs to ask someone else for help. Knowing when to keep working on something yourself, versus asking for help, is a critical soft skill for data science work. 

\subsubsection{Practical ideas for imparting good communication skills}
\begin{itemize}
    \item \textbf{Practice how to ask for help:} Have students practice writing "help" posts and discuss the standard outline of what a call for help should have \citep{stackoverflow}. StackOverflow and other online communities can be very helpful, but often this starts with a well crafted post.
    \item \textbf{Be available:} Always reiterate your availability (and truly be available too). When students do come to you with questions, try to be enthusiastic and supportive of them.    
    \item \textbf{Automate questions:} Set up systems that regularly ask your students what questions or problems they have. You can set up reminders for yourself or them. 
    \item \textbf{Structured one-on-one sessions} Set up structured one-on-one mentoring meetings. By structured, we mean use a document that your student fills out that asks them to answer: what are they working on? what is going well? what is not going so well? and so on.
    \item \textbf{Model good communication yourself:} When live coding, add documentation and try to stick to a code style. Example code should be even more extraordinarily well documented. Emphasize stories where you have messed up code or been stuck and asked someone for help. 
    \item \textbf{Low(er) stakes presentations:} Have students practice presenting to their peers. Public speaking is notoriously scary, particularly if you are presenting on a new topic. The most effective way to make it less scary is to practice. Encourage your students to share their results regularly. When they do present, reaffirm that everyone is rooting for them and no one will interrogate them. Presentations by early professionals are the time to be supportive and enthusiastic, not critique the results or code. 
    \item \textbf{Rubber ducking:} Rubber ducking refers to the debugging code by explaining it aloud, even if no one is listening to your explanation. Encourage students to walk through their code on their own and translate it into "normal speak". This not only helps them troubleshoot, but also builds explanatory skills and deeper understanding of their code. 
\end{itemize}

\section{Iteratively update}

In order to create education that is motivating, inclusive, and realistic, we need to continually update our curriculum and teaching approaches to better serve our students. To best prepare our students for practicing data science, we need to keep pace with the latest data science techniques and educational approaches. This means we should avoid viewing curriculum and teaching methods as set in stone. Instead, we should utilize systems that allow us to easily update and maintain our curriculum and teaching guides \citep{Savonen2023, lau_2022}. Continuous improvement applies in a social context as well; students' interests, career goals, and how they are feeling about the course material is an ongoing conversation.

\subsection{Practical ideas for iteratively updating}

\begin{itemize}
    \item \textbf{Version control your curriculum:} Not only should our data analyses be well tracked, documented and version controlled, but our curriculum should be too. Where possible, curriculum should be open source and on GitHub \citep{Savonen2023}. Also consider using permissive licenses such as a \href{https://creativecommons.org/about/cclicenses}{Creative Commons licenses such as CC-BY} which requires attribution but is otherwise open to repurpose and reuse. 
    \item \textbf{Minimize maintenance pains:} Create your curriculum in a way that minimizes the pain of maintenance. We use Open-source Tools for Training Resources (OTTR) to create our curriculum \citep{Savonen2023}. We also utilize the \href{https://github.com/AlexsLemonade/exrcise}{exercise} package to automatically generate our exercise notebooks without solutions \cite{exrcise}. See the Supplementary info for more resources for how to automate your curriculum maintenance. 
    \item \textbf{Take notes:} In each iteration of your class, take notes and debrief with your education team about strengths of your course and opportunities for improvement. You can easily track ideas and notes as issues in your GitHub repository. 
    \item \textbf{Survey your students:} Use short and focused surveys to take the temperature of the class. Note that some interpretation of surveys are needed. For example, if half your students feel the course speed is "too fast" and the other half feel that the course speed is "too slow", it may mean the course speed is just right. 
\end{itemize}

\section{Limitations}

The tips and philosophies that we have discussed here are quite general. There are many different contexts in which data science may be taught, and many different audiences. It is up to you as the educator to determine which of these ideas would be appropriate. To a certain extent, we also encourage you to experiment with tactics (you are a scientist, after all!) and see what works. Always ask your students how things are working for them. Please comment on this F1000 manuscript or leave us a GitHub issue on a related GitHub repository on one of our affiliated GitHub organizations: \href{https://github.com/jhudsl}{jhudsl} and \href{https://github.com/fhdsl}{fhdsl}.

\section{Conclusion}

Data scientists are in high demand in nearly every modern industry. There is also great potential for using data science in ways that benefit the public good \citep{jones_2023}. The insights and power of data science are exciting, so we feel that the teaching of these skills should be done with a matching level of excitement and with thoughtfulness about the implications of our work. Educational opportunities for learning these skills are not yet equitably distributed but have potential to scale to meet industry's demands for data science. Not only will improving data science education techniques help meet the hiring demand, it will also help empower the lives of young people, researchers, and other students of data science. We have included many tools and resources to help you apply these ideals to your own teaching. We hope that this manuscript opens communication about ways to improve data science teaching approaches in ways that empower everyone in a more equitable manner. Equal opportunities for data science starts with equal opportunities for education.

\section{Future directions}

We the authors will also continue to strive to apply these ideals in our teaching and expand them as we continue to learn. More useful resources for data science teaching are being made every day (see our affiliated GitHub organizations: \href{https://github.com/jhudsl}{jhudsl} and \href{https://github.com/fhdsl}{fhdsl}). We encourage readers to access this article's summary and a longer list of resources discussed here in the Supplementary Materials section. We invite you to contact us, either through GitHub issues or email, if you know of resources that should be added to this list.

\section{Competing Interests}
The authors declare no competing interests. Individual courses we have created on Coursera and Leanpub do generate course revenue, but we do not obtain revenue for any courses other individuals create using OTTR. 

\section{Data availability}
No data is associated with this article.

\section{Grant Information}

This work was supported by the National Cancer Institute under Grant UE5CA254170. AMH, EH, KELC, and FJT were supported by the GDSCN through a contract to Johns Hopkins University (75N92020P00235) and the AnVIL Project through cooperative agreement awards from the National Human Genome Research Institute with cofunding from OD/ODSS to the Broad Institute (U24HG010262) and Johns Hopkins University (U24HG010263). DataTrail is supported by donations from Posit, Bloomberg Philanthropies, the Abell Foundation, and Johns Hopkins Bloomberg School of Public Health. 

\section{Acknowledgements}
We would like to acknowledge all of our colleagues that assisted us with the various projects mentioned who taught us more about teaching data science. We would also like to acknowledge the students we have taught, who continue to inspire us to be better educators and provided valuable and helpful feedback. 

\section{Resource guides}

\begin{itemize}
    \item \textbf{Summary page of practical tips} This page includes all of the ideas and advice in this manuscript into one summarized page as a PDF \href{https://github.com/datatrail-jhu/data_science_teaching_resources/blob/main/README.md}{Link}.
    \item \textbf{Promoting IDARE resources} This page includes links to recommended classes and resources for promoting IDARE principles in the classroom \href{https://github.com/datatrail-jhu/data_science_teaching_resources/blob/main/idare_resources_and_advice.md}{Link}.
    \item \textbf{Curriculum tools} This page includes tools and resources that are very useful for curriculum maintenance and creation \href{https://github.com/datatrail-jhu/data_science_teaching_resources/blob/main/tools_for_curriculum_maintenance.md}{Link}.
    
\end{itemize}

{\small\bibliographystyle{unsrtnat}
\bibliography{references}}

\end{document}